\documentclass[superscriptaddress,showpacs,aps,prd,preprint]{revtex4}
\usepackage{epsfig}
\usepackage{graphicx}
\usepackage{dcolumn}
\usepackage{bm}
\usepackage{url}

\date{}
\begin{document}


\newcommand{\ds}{\displaystyle}
\newcommand{\mc}{\multicolumn}
\newcommand{\bce}{\begin{center}}
\newcommand{\ece}{\end{center}}
\newcommand{\beq}{\begin{equation}}
\newcommand{\eeq}{\end{equation}}
\newcommand{\bea}{\begin{eqnarray}}

\newcommand{\eea}{\end{eqnarray}}
\newcommand{\cont}{\nonumber\eea\bea}
\newcommand{\cl}[1]{\begin{center} {#1} \end{center}}
\newcommand{\ba}{\begin{array}}
\newcommand{\ea}{\end{array}}

\newcommand{\ab}{{\alpha\beta}}
\newcommand{\cd}{{\gamma\delta}}
\newcommand{\dc}{{\delta\gamma}}
\newcommand{\ac}{{\alpha\gamma}}
\newcommand{\bd}{{\beta\delta}}
\newcommand{\abc}{{\alpha\beta\gamma}}
\newcommand{\eps}{{\epsilon}}
\newcommand{\lam}{{\lambda}}
\newcommand{\mn}{{\mu\nu}}
\newcommand{\mpnp}{{\mu'\nu'}}
\newcommand{\Amuu}{{A_{\mu}}}
\newcommand{\Amuo}{{A^{\mu}}}
\newcommand{\Vmuu}{{V_{\mu}}}
\newcommand{\Vmuo}{{V^{\mu}}}
\newcommand{\Anuu}{{A_{\nu}}}
\newcommand{\Anuo}{{A^{\nu}}}
\newcommand{\Vnuu}{{V_{\nu}}}
\newcommand{\Vnuo}{{V^{\nu}}}
\newcommand{\Fmnu}{{F_{\mu\nu}}}
\newcommand{\Fmno}{{F^{\mu\nu}}}

\newcommand{\abcd}{{\alpha\beta\gamma\delta}}


\newcommand{\bsigma}{\mbox{\boldmath $\sigma$}}
\newcommand{\btau}{\mbox{\boldmath $\tau$}}
\newcommand{\brho}{\mbox{\boldmath $\rho$}}
\newcommand{\bpipi}{\mbox{\boldmath $\pi\pi$}}
\newcommand{\bss}{\bsigma\!\cdot\!\bsigma}
\newcommand{\btt}{\btau\!\cdot\!\btau}
\newcommand{\bnabla}{\mbox{\boldmath $\nabla$}}
\newcommand{\bphi}{\mbox{\boldmath $\tau$}}
\newcommand{\bvarphi}{\mbox{\boldmath $\rho$}}
\newcommand{\bDelta}{\mbox{\boldmath $\Delta$}}
\newcommand{\bpsi}{\mbox{\boldmath $\psi$}}
\newcommand{\bPsi}{\mbox{\boldmath $\Psi$}}
\newcommand{\bPhi}{\mbox{\boldmath $\Phi$}}
\newcommand{\bnab}{\mbox{\boldmath $\nabla$}}
\newcommand{\bpi}{\mbox{\boldmath $\pi$}}
\newcommand{\btheta}{\mbox{\boldmath $\theta$}}
\newcommand{\bkappa}{\mbox{\boldmath $\kappa$}}

\newcommand{\bA}{{\bf A}}
\newcommand{\bB}{\mbox{\boldmath $B$}}
\newcommand{\bp}{\mbox{\boldmath $p$}}
\newcommand{\bk}{\mbox{\boldmath $k$}}
\newcommand{\bl}{\mbox{\boldmath $l$}}
\newcommand{\bc}{\mbox{\boldmath $c$}}
\newcommand{\bq}{\mbox{\boldmath $q$}}
\newcommand{\be}{\mbox{\boldmath $e$}}
\newcommand{\bzero}{\mbox{\boldmath $0$}}
\newcommand{\bfe}{{\bf e}}
\newcommand{\bb}{\mbox{\boldmath $b$}}
\newcommand{\br}{\mbox{\boldmath $r$}}
\newcommand{\bR}{\mbox{\boldmath $R$}}

\newcommand{\fph}{${\cal F}$}
\newcommand{\aph}{${\cal A}$}
\newcommand{\dph}{${\cal D}$}
\newcommand{\fpi}{f_\pi}
\newcommand{\mpi}{m_\pi}
\newcommand{\Tr}{{\mbox{\rm Tr}}}
\def\Qb{\overline{Q}}
\newcommand{\delu}{\partial_{\mu}}
\newcommand{\delo}{\partial^{\mu}}
%
%
\newcommand{\up}{\!\uparrow}
\newcommand{\upup}{\uparrow\uparrow}
\newcommand{\updo}{\uparrow\downarrow}
\newcommand{\uu}{$\uparrow\uparrow$}
\newcommand{\ud}{$\uparrow\downarrow$}
\newcommand{\auu}{$a^{\uparrow\uparrow}$}
\newcommand{\aud}{$a^{\uparrow\downarrow}$}
\newcommand{\pu}{p\!\uparrow}

\newcommand{\qp}{quasiparticle}
\newcommand{\sa}{scattering amplitude}
\newcommand{\ph}{particle-hole}
\newcommand{\qcd}{{\it QCD}}
\newcommand{\integ}{\int\!d}
\newcommand{\ie}{{\sl i.e.~}}
\newcommand{\etal}{{\sl et al.~}}
\newcommand{\etc}{{\sl etc.~}}
\newcommand{\rhs}{{\sl rhs~}}
\newcommand{\lhs}{{\sl lhs~}}
\newcommand{\eg}{{\sl e.g.~}}
\newcommand{\ef}{\epsilon_F}
\newcommand{\sigt}{d^2\sigma/d\Omega dE}
\newcommand{\sige}{{d^2\sigma\over d\Omega dE}}
\newcommand{\rpaeq}{\beq
\left ( \begin{array}{cc}
A&B\\
-B^*&-A^*\end{array}\right )
\left ( \begin{array}{c}
X^{(\kappa})\\Y^{(\kappa)}\end{array}\right )=E_\kappa
\left ( \begin{array}{c}
X^{(\kappa})\\Y^{(\kappa)}\end{array}\right )
\eeq}
\newcommand{\ket}[1]{| {#1} \rangle}
\newcommand{\bra}[1]{\langle {#1} |}
\newcommand{\ave}[1]{\langle {#1} \rangle}
\newcommand{\half}{{1\over 2}}

\newcommand{\singlespace}{
    \renewcommand{\baselinestretch}{1}\large\normalsize}
\newcommand{\doublespace}{
    \renewcommand{\baselinestretch}{1.6}\large\normalsize}
\newcommand{\bftau}{\mbox{\boldmath $\tau$}}
\newcommand{\bfalpha}{\mbox{\boldmath $\alpha$}}
\newcommand{\bfgamma}{\mbox{\boldmath $\gamma$}}
\newcommand{\bfxi}{\mbox{\boldmath $\xi$}}
\newcommand{\bfbeta}{\mbox{\boldmath $\beta$}}
\newcommand{\bfeta}{\mbox{\boldmath $\eta$}}
\newcommand{\bfpi}{\mbox{\boldmath $\pi$}}
\newcommand{\bfphi}{\mbox{\boldmath $\phi$}}
\newcommand{\bfR}{\mbox{\boldmath ${\cal R}$}}
\newcommand{\bfL}{\mbox{\boldmath ${\cal L}$}}
\newcommand{\bfM}{\mbox{\boldmath ${\cal M}$}}
\newcommand{\aem}{\mbox{$\alpha_{\rm{em}}$}}
\newcommand{\Lmn}{\mbox{$L_{\mu \nu}$}}
\newcommand{\Wmn}{\mbox{$W_{\mu \nu}$}}
\newcommand{\uLmn}{\mbox{$L^{\mu \nu}$}}
\newcommand{\uWmn}{\mbox{$W^{\mu \nu}$}}

\def\dblint{\mathop{\rlap{\hbox{$\displaystyle\!\int\!\!\!\!\!\int$}}
    \hbox{$\bigcirc$}}}
\def\ut#1{$\underline{\smash{\vphantom{y}\hbox{#1}}}$}

\def\UNITY{{\bf 1\! |}}
\def\Pom{{\bf I\!P}}
\def\lsim{\mathrel{\rlap{\lower4pt\hbox{\hskip1pt$\sim$}}
    \raise1pt\hbox{$<$}}}         
\def\gsim{\mathrel{\rlap{\lower4pt\hbox{\hskip1pt$\sim$}}
    \raise1pt\hbox{$>$}}}         
\def\beq{\begin{equation}}
\def\eeq{\end{equation}}
\def\bea{\begin{eqnarray}}
\def\eea{\end{eqnarray}}

        
\title{Low mass Drell-Yan production of lepton pairs \\ at forward
  directions at the LHC: a hybrid approach}

\author{Wolfgang Sch\"afer}%
\email{Wolfgang.Schafer@ifj.edu.pl}
\affiliation{Institute of Nuclear Physics PAN, PL-31-342 Cracow, Poland}

\author{Antoni Szczurek%
\footnote{Also at University of Rzesz\'ow, PL-35-959 Rzesz\'ow, Poland.}}
\email{Antoni.Szczurek@ifj.edu.pl}
\affiliation{Institute of Nuclear Physics PAN, PL-31-342 Cracow, Poland}

\begin{abstract}
We discuss Drell-Yan production of dileptons at high energies 
in forward rapidity region in a hybrid high-energy approach. This approach 
uses unintegrated gluon distributions in one proton and collinear
quark/antiquark distributions in the second proton.
Corresponding momentum-space formula for the differential cross sections 
in high-energy approximation has been derived and will be presented.
The relation to the commonly used dipole approach is discussed.
We conclude and illustrate that some results of the dipole approaches 
are too approximate, as far as kinematics is considered, and in fact 
cannot be used when comparing with real experimental data.
We find that the dipole formula is valid only in very forward/backward
rapidity regions ($|y| >$ 5) that cannot be studied experimentally
in the moment.
We performed calculations of some differential cross sections for
low-mass dilepton production by the LHCb and ATLAS collaborations.
In distinction to most of dipole approaches, we include all of the four
Drell-Yan structure functions, although the impact of interference
structure functions is rather small for the relevant experimental
cuts.
We find that both side contributions ($g + q/\bar q$ and $q/\bar q + g$) 
have to be included even for the LHCb rapidity coverage which
is in contradiction with what is usually done in the dipole approach.
We present results for different unintegrated gluon distributions from
the literature (some of them include saturation effects).
We see no clear hints of saturation even at small $M_{ll}$
when comparing with the LHCb data.
\end{abstract}
\date{\today}
\maketitle



\section{Introduction}

The Drell-Yan process of inclusive lepton-pair production is
one of the important sources on the partonic structure of protons
\cite{Drell:1970wh,Ellis:1991qj,Peng:2014hta}.
It was proposed some time ago that the Drell-Yan production of low
invariant masses of dileptons in forward directions could be another 
good place in searching for the onset of (gluon) saturation 
\cite{Gelis:2002fw,Gelis:2006hy}.
A number of different approaches have recently been used to calculate
Drell-Yan processes in the small-$x$ region.

In particular in recent applications for LHC much attention 
has been paid to the color-dipole approach 
\cite{Ducati:2013cga,GolecBiernat:2010de,Basso:2015pba,Motyka:2014lya},
in which the main ingredient is 
the color dipole-nucleon cross section \cite{Nikolaev:1990ja}
parametrized as a function of dipole size  and collision energy
or a similar equivalent kinematical variable. 

Alternatively a $k_T$-factorization approach is used to describe
dilepton production. Here the recent works
 \cite{Szczurek:2008ga,Nefedov:2012cq,Baranov:2014ewa}
are based on quark and antiquark unintegrated distributions. 
This formulation however is not adequate to address the
nonlinear effects in the gluon distribution dubbed ``saturation''.   
Another approach relates the small-$x$ unintegrated quark density
explicitly to the unintegrated gluon distribution \cite{Hautmann:2012sh}.

Most of the above calculations, especially in the color-dipole framework 
do not address lepton momentum and angular distributions, 
but rather concentrate on a few observables, such as the  
dilepton invariant mass, rapidity and transverse momentum.
All of these observables can be expressed through the 
inclusive production cross sections of a virtual heavy photon,
which carries either transverse or longitudinal polarization.

For the full description of lepton distributions this is however not enough--
there are interferences between transverse and longitudinal and different
transverse polarization to be taken into account. The complete description
of the Drell-Yan process therefore requires four structure functions 
\cite{Oakes:1966,Lam:1978pu,Boer:2006eq}.

\begin{figure}[!ht]
\includegraphics[width=5cm]{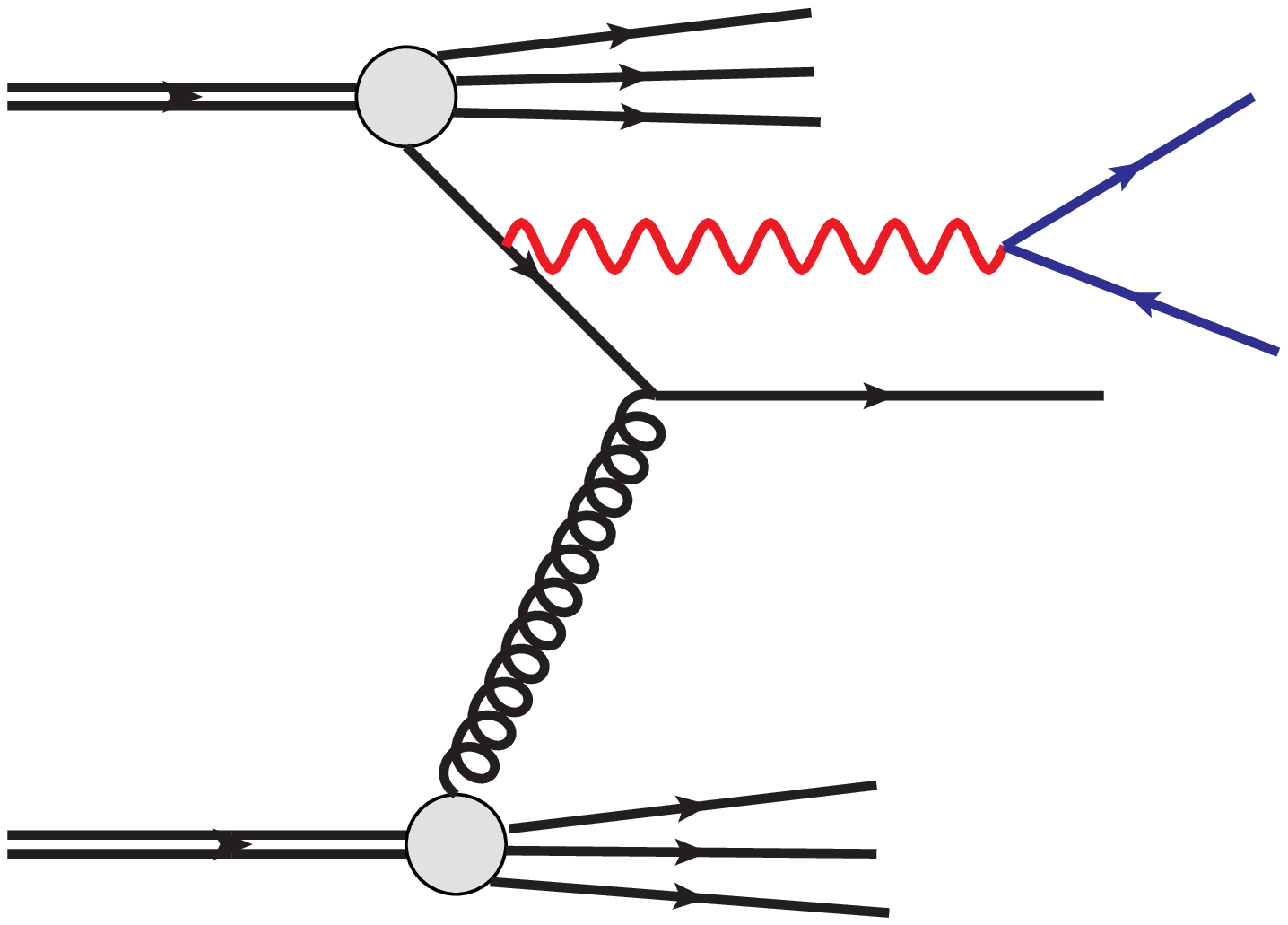}
\includegraphics[width=5cm]{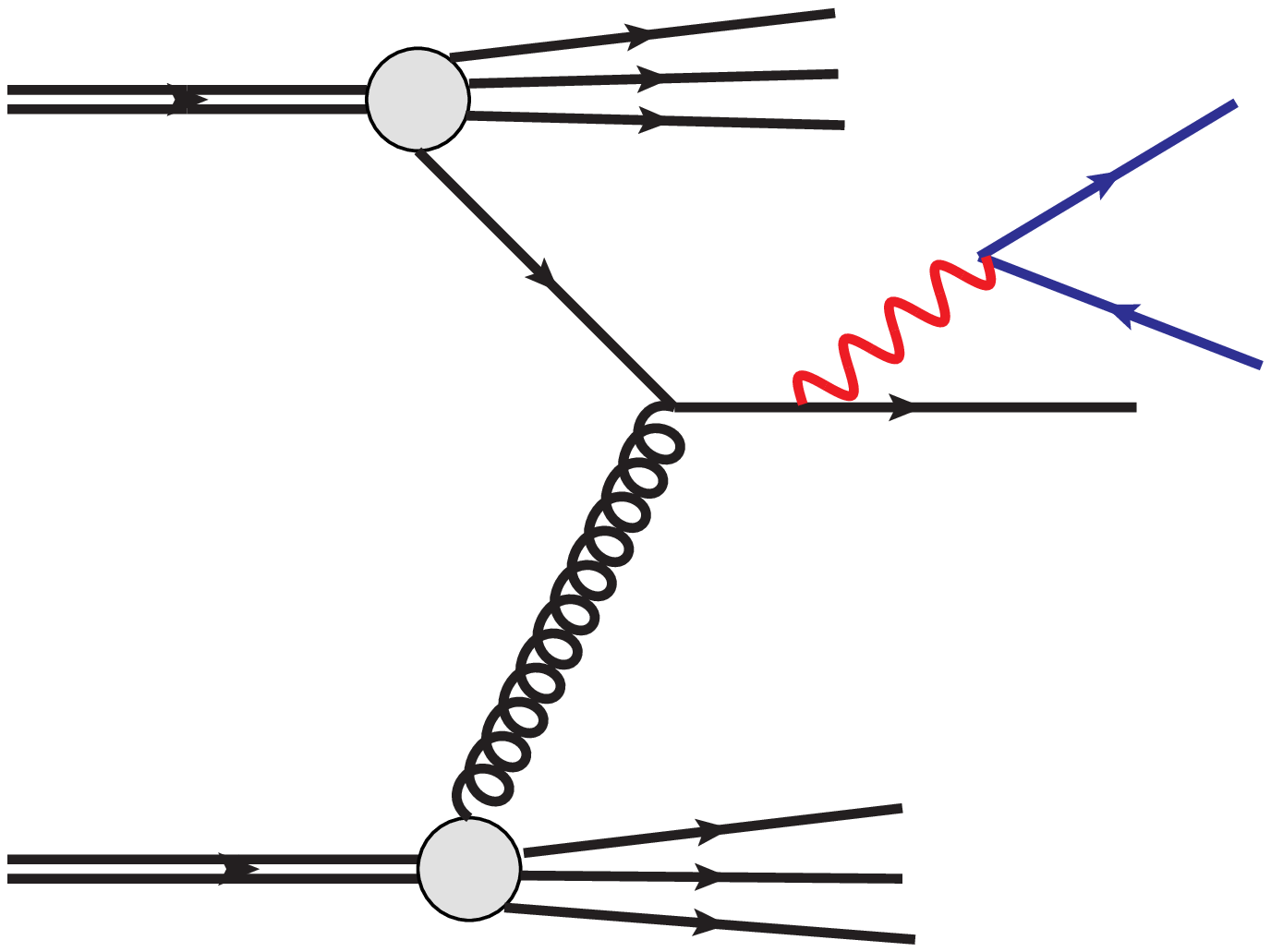}\\
\includegraphics[width=5cm]{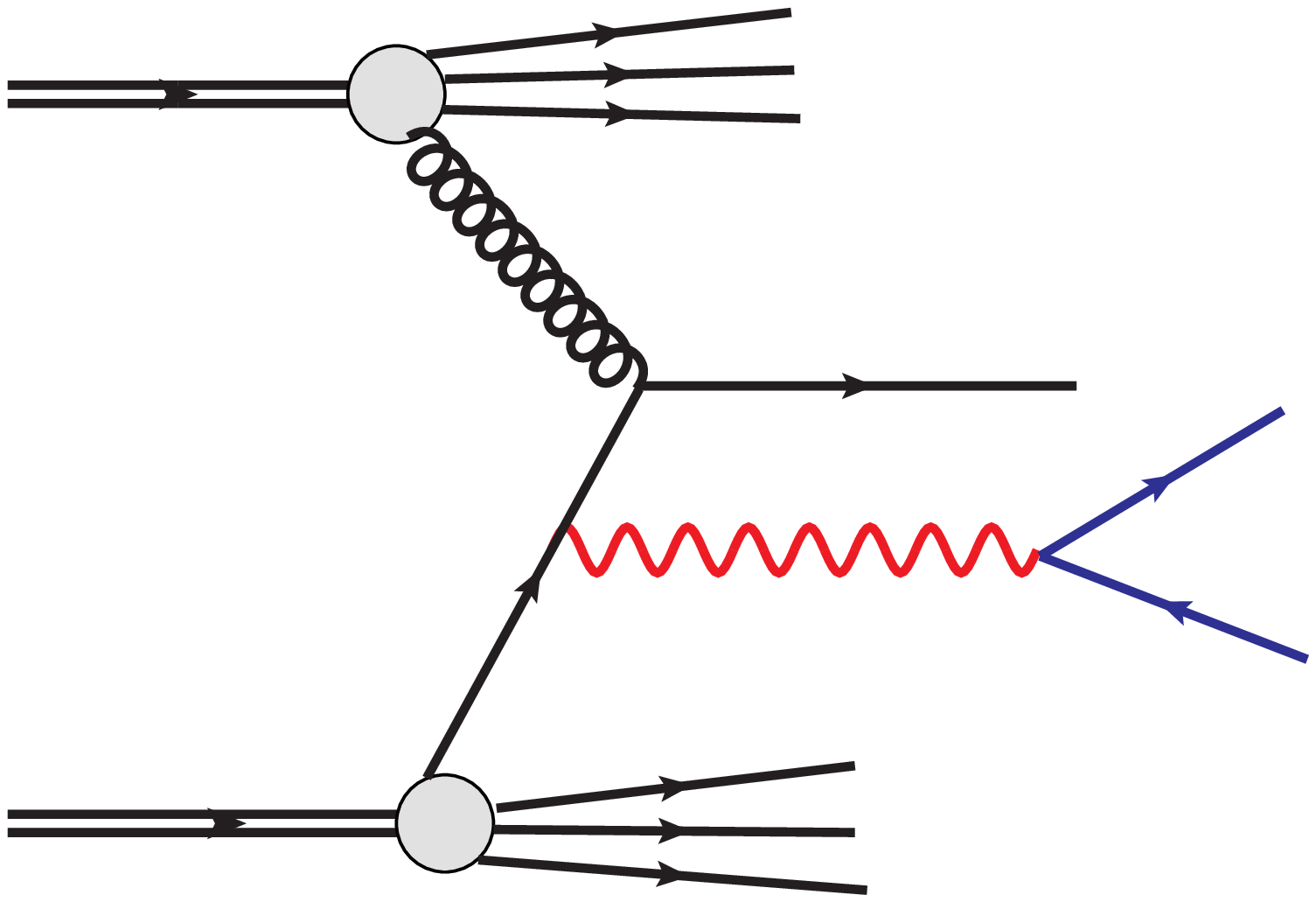}
\includegraphics[width=5cm]{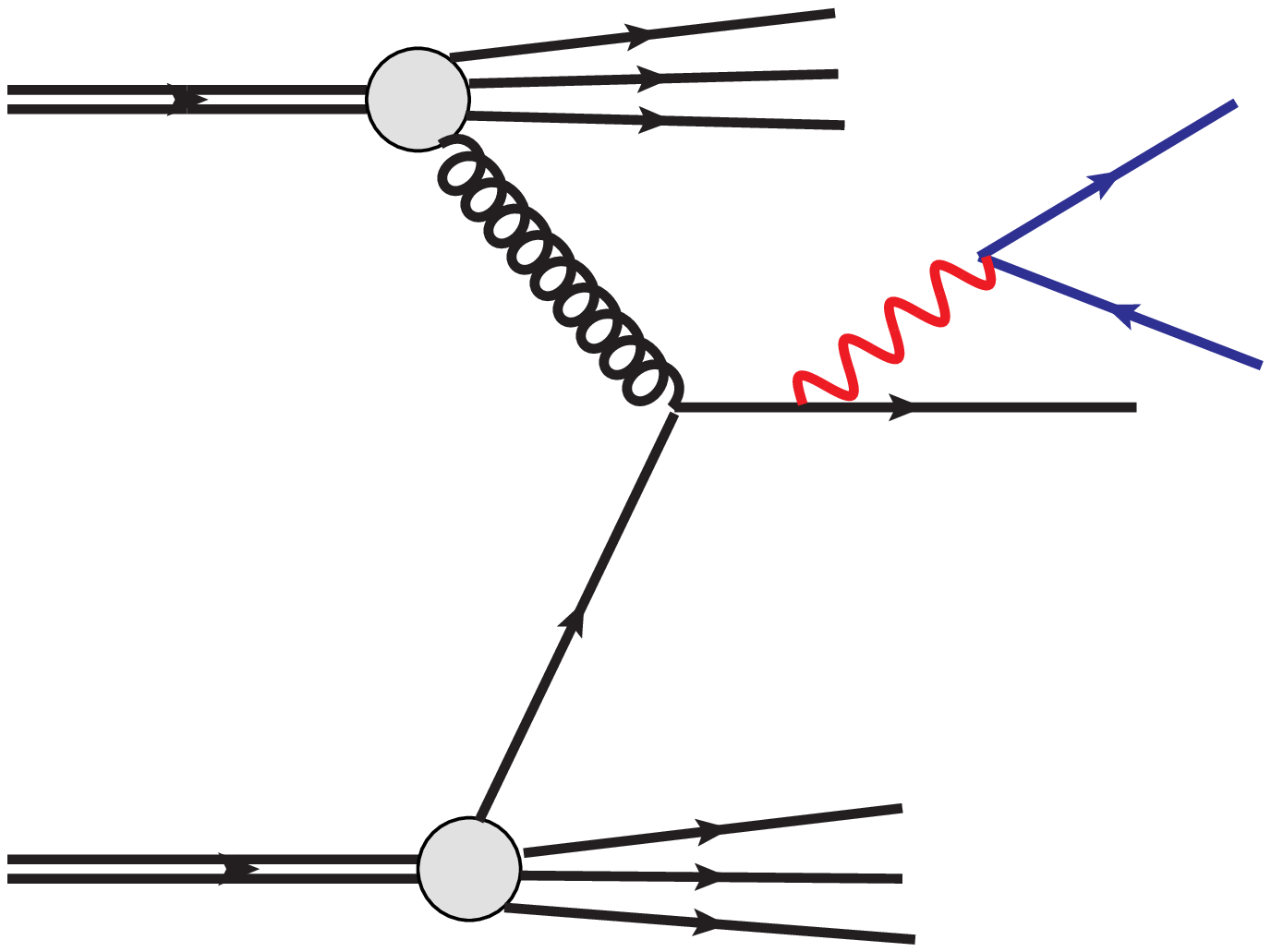}
\caption{\label{fig:diagrams}
The diagrams relevant for forward and backward production of dilepton pairs.
}
\end{figure}

In this paper we shall also start from the impact parameter 
representation, but we will perform the Fourier transformation to transverse
momentum space. What then emerges \cite{Gelis:2002fw} is a 
hybrid collinear/$k_T$-factorization, in which  
the main ingredients will be collinear quark/antiquark and 
unintegrated gluon distributions (see e.g. \cite{Deak:2009xt} for predictions
of forward jets in such an approach). The dominant processes captured by this approach
are shown in Fig.\ref{fig:diagrams}.
The present approach allows for explicit treatment and control 
of momenta of individual leptons ($e^+ e^-$ or $\mu^+ \mu^-$) and 
therefore a comparison to existing experimental data.
Below, we will also use unintegrated gluon distribution functions (UGDFs) equivalent 
to the dipole-nucleon cross sections known from the literature.
Then a direct comparison of results from different dipole models/UGDFs 
with experimental data \cite{LHCb-CONF_2012,Chatrchyan:2013tia,Aad:2014qja} will be possible.

\section{Inclusive lepton pair production: kinematics, frames, structure functions}

The cross section for inclusive $l^+ l^-$ production (Drell-Yan process) can be presented as
\begin{eqnarray}
(2 \pi)^4 {d \sigma (pp \to l^+(k_+) l^-(k_-) X)  \over d^4 q} = { (4 \pi \aem)^2 \over 2 S M^4}  \cdot \Wmn L^{\mu \nu}  \cdot  d\Phi(q,k_+,k_-) \, .
\end{eqnarray}
Here $q = k_+ + k_-$ is the four-momentum of the virtual photon, $q^2 = M^2$ is the invariant mass of the lepton pair.
The lepton-tensor $\Lmn$, is known explicitly:
\begin{eqnarray}
 \Lmn = 4 \cdot \Big( k_{+ \mu} k_{- \nu} + k_{- \mu} k_{+ \nu} - {M^2 \over 2} g_{\mu\nu} \Big) \, . 
\end{eqnarray}
All dynamical information on the production process of the virtual photon is contained in the hadronic
tensor
\begin{eqnarray}
 \Wmn =  \int d^4x \exp(-i q\cdot x) \,  \bra{p_1 p_2} J^{\rm{em}}_\mu(0) J^{\rm{em}}_\nu(x) \ket{p_1 p_2} \, .
\end{eqnarray}
One conventionally decomposes the hadronic tensor introducing four structure functions
\cite{Oakes:1966,Lam:1978pu}:
\begin{eqnarray}
 \Wmn = (\hat x_\mu \hat x_\nu + \hat y_\mu \hat y_\nu) W_T + \hat z _\mu \hat z_\nu W_L  + (\hat y_\mu \hat y _\nu - \hat x_\mu \hat x_\nu) W_{\Delta \Delta}
 - (\hat x_\mu \hat z_\nu + \hat z_\mu \hat x_\nu) W_\Delta \, ,
\end{eqnarray}
where the covariant directions $\hat x_\mu, \hat y_\mu, \hat z_\mu$ define the spatial axes
in a rest frame of the dilepton pair (or the massive photon).

The individual structure functions can be projected out by contraction with helicity states of the massive photon as
\begin{eqnarray}
W_T &=& \uWmn \epsilon^{(+)}_\mu \epsilon^{(+)*}_\nu, \, W_L = \uWmn \epsilon^{(0)}_\mu \epsilon^{(0)}_\nu \, , \nonumber \\
W_{\Delta} &=& \uWmn (\epsilon_\mu^{(+)} \epsilon_\nu^{(0)} + \epsilon_\mu^{(0)} \epsilon^{(+)*}_\nu ) {1 \over \sqrt{2}} \, , 
W_{\Delta \Delta} = \uWmn \epsilon^{(+)}_\mu \epsilon^{(-)*}_\nu \, . \nonumber \\
\end{eqnarray}
Here, the helicity states are defined as
\begin{eqnarray}
 \epsilon^{(\pm)}_\mu = -{1 \over \sqrt{2}} (\pm \hat x_\mu + i \hat y_\mu)\, , \epsilon^{(0)}_\mu = \hat z_\mu \, .
\end{eqnarray}
It is furthermore useful to introduce the time direction
\begin{eqnarray}
 \hat t_\mu = {1 \over M} q_\mu \, , \hat t^2 = + 1 \, ,
\end{eqnarray}
and the ``spatial unit matrix''
\begin{eqnarray}
 - \tilde g_{\mu \nu} \equiv \hat x_\mu \hat x_\nu + \hat y_\mu  \hat y_\nu + \hat z_\mu \hat z_\nu  = - g_{\mu \nu} + \hat t_\mu \hat t_\nu = - g_{\mu \nu} + {q_\mu q_\nu \over M^2} \, , 
\end{eqnarray}
To fully define the frame, we should relate the vectors $\hat x_\mu, \hat y_\mu, \hat z_\mu$ to the momenta of measured particles.
From now on, we will use a dilepton rest-frame, in which the $z$-axis points along the momentum of one of the incoming protons (we choose the momentum $p_2$) in that frame.
Such a frame is often called a Gottfried-Jackson frame. For a useful discussion of different frame choices, see \cite{Boer:2006eq}.
Explicitly, we have
\begin{eqnarray}
 &&\hat z_\mu = {M \over q \cdot p_2} \Big( p_{2 \mu} - {q \cdot p_2 \over M^2} q_\mu \Big) = { M \over q \cdot p_2} \tilde p_{2\mu} \, , \nonumber \\
 &&\hat x_\mu = {\sqrt{-q^2_\perp} \over q \cdot p_2} \Big(p_{2 \mu} - { q \cdot p_2  \over q_\perp^2 } q_{\perp \mu} \Big) = { M \over \sqrt{-q_\perp^2} (p_1 \cdot p_2)} \Big( (p_2\cdot \hat z) \tilde p_{1\mu} 
- (p_1\cdot \hat z) \tilde p_{2\mu} \Big) \, , \nonumber \\
 &&\hat y_\mu = \varepsilon_{\mu \alpha \beta \gamma} \hat x_\alpha \hat z_\beta \hat t_\gamma = {1 \over \sqrt{-q_\perp^2}} \, \varepsilon_{\mu \alpha \beta \gamma} n^+_\alpha n^-_\beta q_{\perp \gamma} \, . 
\end{eqnarray}
Here we used the notation 
\begin{eqnarray}
 \tilde p_{i \mu} \equiv \tilde g_{\mu \nu} p_{i \nu} = p_{i \mu} - {(p_i \cdot q) \over M^2} q_\mu \, , i = 1,2 \, ,
\end{eqnarray}
as well as
\begin{eqnarray}
 n^+_\mu &=& \sqrt{2 \over S} \, p_{1\mu} \, , \, n^-_\mu = \sqrt{2 \over S} \, p_{2 \mu} \, , \nonumber \\
 q_{\perp \mu} &=& \Big( g_{\mu \nu} - n^+_\mu n^-_\nu - n^-_\mu n^+_\nu \Big) q^\nu \, .
\end{eqnarray}
Notice, that $q_{\perp \mu}$  is the transverse momentum of the virtual photon in the $p p$-center of mass frame.
Below, boldface letters will denote the two-dimensional transverse momenta, so that e.g. $q_\perp^2 = - \bq^2$. 
We will also use the notation $q_T \equiv |\bq|$ for the absolute values of two-dimensional vectors.

Now, performing explicitly the contraction of leptonic and hadronic tensor expressed in the chosen basis, 
we obtain the inclusive dilepton cross section as
\begin{eqnarray}
{d\sigma(pp \to l^+ l^- X) \over dx_+ dx_- d^2\bk_+ d^2\bk_-} = &&{ \aem \over (2 \pi)^2 M^2} {x_F \over x_+ x_-} \Big\{ \Sigma_T(x_F,\bq,M^2)  D_T\Big({x_+ \over x_F} \Big)  +
\Sigma_L(x_F,\bq,M^2)  D_L\Big({x_+ \over x_F}\Big) \nonumber \\
&& + \Sigma_\Delta(x_F,\bq,M^2)  D_\Delta\Big({x_+ \over x_F}\Big)  \Big({\bl \over |\bl|} \cdot {\bq \over |\bq|} \Big) \nonumber \\
&& +  \Sigma_{\Delta \Delta}(x_F,\bq,M^2) D_{\Delta \Delta}\Big({x_+
  \over x_F}\Big) \Big(2  \Big({\bl \over |\bl|} \cdot {\bq \over |\bq|}
\Big)^2 - 1 \Big) \Big\} \; .
\end{eqnarray}
We use the light-cone parametrization of particle momenta:
\begin{eqnarray}
 k^{\pm}_{\mu} &=& x_\pm \sqrt{S\over 2} \, n^+_\mu +  {\bk_\pm^2 \over x_\pm \sqrt{2 S}} \, n^-_\mu + k^\pm_{\perp \mu} \, , \nonumber \\
 q_\mu &=& x_F \sqrt{S \over 2} \, n^+_\mu +   {M^2 + \bq^2 \over x_F \sqrt{2S}} \, n^-_\mu + q_{\perp \mu} \, .
\end{eqnarray}
so that 
\begin{eqnarray}
 x_F = x_+ + x_- , \bq = \bk_+ + \bk_- \, .
\end{eqnarray}
We also need the light-cone relative transverse momentum
\begin{eqnarray}
 \bl = {x_+ \over x_F} \bk_- - {x_- \over x_F} \bk_+ \, .
\end{eqnarray}
The functions $D_i, i \in \{T,L,\Delta,\Delta \Delta\}$ come from the contractions of the leptonic tensor and describe the
$\gamma^* \to l^+ l^-$ transition. They are given by
\begin{eqnarray}
  D_T(u) = 4 \Big( u^2 + (1-u)^2 \Big) \; ,\nonumber \\
 D_L(u) = D_{\Delta \Delta}(u) = 8 u (1-u) \; , \nonumber \\
 D_\Delta(u) = 4 \sqrt{u(1-u)} \, (2u-1) \, . 
\end{eqnarray}
Finally, the functions $\Sigma_i(x_F,\bq,M^2), i \in \{T,L,\Delta,\Delta \Delta\}$ parametrize the density matrix
of production of the massive photon.
Expressed in terms of helicity eigenstates, we have for the density matrix
\begin{eqnarray}
\rho_{\lambda \lambda'} {d \sigma (pp \to \gamma^*(M^2) X) \over dx_F d^2\bq} = {1 \over x_F} {\aem \over 8 \pi^2 S} \Wmn \epsilon_\mu^{(\lambda)} \epsilon_\nu^{(\lambda')*}  \, .
\end{eqnarray}
Or
\begin{eqnarray}
 \rho_{\lambda \lambda'} = {\Wmn  \epsilon_\mu^{(\lambda)} \epsilon_\nu^{(\lambda')*}  \over 2 W_T + W_L} \, , \,  \rho_{++} + \rho_{--} + \rho_{00} = 1 \, .
\end{eqnarray}
Above we used the components
\begin{eqnarray}
 \Sigma_i(x_F,\bq,M^2) = \rho_i {d \sigma (pp \to \gamma^*(M^2) X) \over dx_F d^2\bq} \equiv {1 \over x_F} {\aem \over 8 \pi^2 S} W_i, \, \, i \in{T,L, \Delta,\Delta \Delta} \, .
\end{eqnarray}

\section{The parton level process: $q p \to \gamma^* X$}

Let us now turn to the parton-level description of the Drell-Yan process. 
What we ultimately need are the hadron-level density matrix elements for the
$pp \to \gamma^* X$ process.
As we are interested in the ``forward region'' of phase space, it is reasonable to assume
that the most important degrees of freedom will be quarks and antiquarks from one of the protons
and small-$x$ gluons from the second one.
Our parton level subprocess will therefore look like an excitation of the $\gamma^* q$-Fock state of
an incoming quark in the small-$x$ gluon field of the second hadron.

We follow the notation and normalization of \cite{Nikolaev:2004cu}, and can write down the 
density-matrix for production of the virtual photon in the $q p \to \gamma^* X$ process as
\begin{eqnarray}
 \hat \rho_{\lambda \lambda'} {d \hat \sigma (qp \to \gamma^*(z,\bq) X) \over dz d^2\bq} = {1 \over 2 (2 \pi)^2} &&   \overline{\sum_{\sigma, \sigma'}}  \int d^2\br d^2 \br' \exp[-i \bq(\br-\br')] 
\psi_{\sigma \sigma'}^{(\lambda)}(z,\br) \psi_{\sigma \sigma'}^{(\lambda')*}(z,\br') \nonumber \\
&&\times \Big( \sigma(x_2,z \br ) + \sigma(x_2, z \br') - \sigma (x_2, z (\br - \br')) \Big) \, .
\label{eq:parton-level}
\end{eqnarray}
The light-front wave functions for the $q_\sigma \to \gamma^*_\lambda q_\sigma'$ transition
(here $\sigma, \sigma', \lambda$ denote the helicities of particles) read:
\begin{eqnarray}
\psi_{\sigma \sigma'}^{(\lambda)}(z,\br) &=& \int {d^2 \bq \over (2 \pi)^2} \exp[-i \br \bq]  \psi_{\sigma \sigma'}^{(\lambda)}(z,\bq) \nonumber \\
&=& e_q \sqrt{z (1-z)} \int  {d^2 \bq \over (2 \pi)^2} \exp[-i \br \bq] { \bar{u}_{\sigma'} (1-z,-\bq) \epsilon^{(\lambda)*}_\mu \gamma_\mu u_\sigma(1,\bzero) 
\over \bq^2 + \varepsilon^2},
\label{eq:WF}
\end{eqnarray}
with $\varepsilon^2 = (1-z) M^2 + z^2 m_q^2$.

To derive the momentum-space $k_T$-factorization representation, we use the relation of the dipole cross section
with the unintegrated gluon distribution
\begin{eqnarray}
 \sigma(x,\br) = { 1 \over 2} \int d^2\bkappa \, f(x,\bkappa) ( 1 - \exp[i\bkappa \br]) ( 1 - \exp[-i \bkappa \br]) \, . 
\label{eq:dip-xsec}
\end{eqnarray}
Where $f(x,\bkappa)$ is
\begin{eqnarray}
 f(x,\bkappa) = {4 \pi  \alpha_S \over N_c} {1 \over \bkappa^4} {\partial G(x,\bkappa^2) \over \partial \log \bkappa^2} \, . 
\end{eqnarray}
Inserting (\ref{eq:WF}) and (\ref{eq:dip-xsec}) into
Eq. (\ref{eq:dip-xsec}), we obtain:
\begin{eqnarray}
 \hat \rho_{\lambda \lambda'} {d \hat \sigma (qp \to \gamma^*(z,\bq) X) \over dz d^2\bq} = {1 \over 2 (2 \pi)^2} \overline{\sum_{\sigma, \sigma'}} \int d^2 \bkappa f(x_2,\bkappa)  \nonumber \\
 \Big( \psi_{\sigma \sigma'}^{(\lambda)}(z,\bq)  - \psi_{\sigma \sigma'}^{(\lambda)}(z,\bq - z\bkappa) \Big) \Big(\psi_{\sigma \sigma'}^{(\lambda')}(z,\bq) -  
\psi_{\sigma \sigma'}^{(\lambda')}(z,\bq - z \bkappa) \Big)^*   
\end{eqnarray} 
From here,
we obtain the impact-factor representation for the elements of 
the density matrix of production $\Sigma_i$, where 
$i = T,L,\Delta,\Delta \Delta$, 
\begin{eqnarray}
 && \hat \Sigma_i(z,\bq,M^2) = \hat \rho_i  {d \hat \sigma (qp \to \gamma^*(z,\bq) X) \over dz d^2\bq} = {e_q^2 \aem \over 2 N_c} \int {d^2\bkappa \over \pi \bkappa^4} \, 
\alpha_{\rm{S}}(\bar q^2) {\cal{F}}(x_2, \bkappa^2) \, I_i(z,\bq,\bkappa) \, ,
\nonumber \\
\label{eq:k_T-fact}
\end{eqnarray}
with
\begin{eqnarray}
&& I_T(z,\bq,\bkappa)= {1 + (1-z)^2 \over z}|\bPhi|^2 
+ z^3 m_q^2 \Phi_0^2  \, , \nonumber \\
&& I_L(z,\bq,\bkappa) = {4 (1-z)^2 M^2 \over z} \, \Phi^2_0  \, ,\nonumber \\
&& I_\Delta(z,\bq,\bkappa) = {2 (2-z)(1-z) M\over z} \Big({\bq \over |\bq|} \cdot \bPhi\Big) \Phi_0 \, , \nonumber \\
&& I_{\Delta \Delta}(z,\bq,\bkappa) = {2( 1-z )\over z} \Big( |\bPhi|^2  - 2 \Big( {\bq \over |\bq|} \cdot \bPhi \Big)^2 \Big) \, ,
\end{eqnarray}
where
\begin{eqnarray}
\bPhi(z,\bq,\bkappa) = {\bq \over \bq^2 + \varepsilon^2} - { \bq -
  z\bkappa \over (\bq - z \bkappa)^2 + \varepsilon^2} \, , \nonumber \\
\Phi_0(z,\bq,\bkappa) =  {1 \over \bq^2 + \varepsilon^2} - {1 \over (\bq - z \bkappa)^2 + \varepsilon^2} \, .
\end{eqnarray}

A brief comment on our $k_T$-factorization form of the Drell-Yan cross section is in order.  
An important property of Eq.(\ref{eq:k_T-fact}) is its linear dependence of the unintegrated glue.
This linear dependence remains valid even in the presence of multiple scattering effects which
become important in the presence of a large saturation scale.  
In fact all possible saturation effects get absorbed into the nonlinear evolution \cite{Balitsky:1995ub,Kovchegov:1999yj} 
of the  unintegrated gluon distribution.

The origin of this simplification is the fact, that the emitted photon does not couple 
to the exchanged gluon \cite{Nikolaev:2005qs}. Indeed for 
the analogous $q \to qg$ transition relevant to the production of forward jets, the linear 
$k_T$-factorization is strongly violated, and the relevant saturation effects are not exhausted
by the nonlinear evolution of the unintegrated glue \cite{Nikolaev:2005dd}.

In a language, where interactions of the fast quark with the target is described by the
correlators of Wilson lines, see e.g. \cite{Dominguez:2011wm}, the above simplification
manifests itself through the fact that the cross section depends only on the correlator
of two fundamental Wilson lines. Higher order correlation functions, which
would have their own evolution equations \cite{Iancu:2011ns}, do not appear.

Therefore there is a sound theoretical motivation behind the search for saturation
effects on the unintegrated glue by means of the forward Drell-Yan process.

\section{$k_T$-factorization form of the dilepton cross section at the hadron level}

To go to the hadron level, we will assume the collinear factorization on the quark side and write, choosing a factorization scale
$\mu^2 \sim \bq^2 + \varepsilon^2$:
\begin{eqnarray}
\Sigma_i (x_F,\bq,M) &=& \sum_f \int dx_1 dz \, \delta(x_F - z x_1) \,  \Big[q_f(x_1,\mu^2) + \bar q_f(x_1,\mu^2 )\Big] \hat \Sigma_i(z,\bq,M^2) \, .   \nonumber \\
&=& \sum_f  {e_f^2 \alpha_{\rm em} \over 2 N_c} \int_{x_F}^1  dx_1 \, \Big[q_f(x_1,\mu^2) + \bar q_f(x_1,\mu^2 )\Big] \int {d^2 \bkappa_2 \over \pi \bkappa_2^4} 
{\cal{F}}(x_2,\bkappa_2^2) 
\alpha_S(\bar q^2) I_i \Big( {x_F \over x_1} ,\bq,\bkappa_2 \Big) \, . \nonumber \\
\end{eqnarray}
The full dilepton cross section is then 
\begin{eqnarray}
{d \sigma (pp \to l^+ l^- X)\over dy_+ dy_- d^2\bk_+ d^2\bk_-} &=& x_+ x_- {d \sigma (pp \to l^+ l^- X) \over dx_+ dx_- d^2\bk_+ d^2\bk_-} \nonumber \\
&=&{\alpha^2_{\rm em} \over 8 \pi^2 N_c M^2} \sum_f e_f^2 
 \int_{x_F}^1  dx_1 \, \Big[ x_1 q_f(x_1,\mu^2) + x_1 \bar q_f(x_1,\mu^2 )\Big] \int {d^2 \bkappa_2 \over \pi \bkappa_2^4} 
{\cal{F}}(x_2,\bkappa_2^2) 
\alpha_S(\bar q^2) \nonumber \\
&& \Big\{ {x_F \over x_1} I_T \Big( {x_F \over x_1} ,\bq,\bkappa_2 \Big) \, D_T\Big({x_+ \over x_F}\Big) + {x_F \over x_1}  I_L \Big( {x_F \over x_1} ,\bq,\bkappa_2 \Big) \, 
D_L\Big({x_+ \over x_F}\Big)  \, \nonumber \\
&& +  {x_F \over x_1} I_\Delta \Big( {x_F \over x_1} ,\bq,\bkappa_2 \Big)  \,
 D_\Delta\Big({x_+ \over x_F}\Big)  \Big({\bl \over |\bl|} \cdot {\bq \over |\bq|} \Big)  \nonumber \\
&&+  {x_F \over x_1} I_{\Delta \Delta} \Big( {x_F \over x_1} ,\bq,\bkappa_2 \Big) 
D_{\Delta \Delta}\Big({x_+ \over x_F}\Big) \Big( 2 \Big({\bl \over
  |\bl|} \cdot {\bq \over |\bq|} \Big)^2 - 1) \Big\} \; .
\end{eqnarray}

If we also want to include the recoiling jet, we can do this by inserting the delta-functions
\begin{eqnarray}
 dx_J \delta(x_J + x_F - x_1) \, d^2\bk_J \, \delta^{(2)}(\bkappa_2 - \bq - \bk_J) \, .
\end{eqnarray}
This gives us the fully differential spectrum
\begin{eqnarray}
 {d \sigma (pp \to l^+ l^- X)\over dy_+ dy_- dy_J  d^2\bk_+ d^2\bk_- d^2\bk_J}   &=& {\alpha^2_{\rm em} \over 8 \pi^3 N_c M^2} {x_F x_J \over x_F + x_J} \nonumber \\
&& \times \sum_f e_f^2 
 \Big[  q_f(x_F + x_J, \mu^2) + \bar q_f(x_F + x_J, \mu^2) \Big] { \alpha_S(\bar q^2) {\cal{F}}(x_2, \bq + \bk_J) \over  (\bq + \bk_J)^4} 
\nonumber \\
&& \times 
\Big\{  I^f_T \Big( {x_F \over x_F + x_J } ,\bq,\bq + \bk_J \Big) \, D_T\Big({x_+ \over x_F}\Big) \nonumber \\
&+&  I^f_L \Big( {x_F \over x_F + x_J} ,\bq,\bq + \bk_J \Big) \, 
D_L\Big({x_+ \over x_F}\Big)  \, \nonumber \\
&+&  I^f_\Delta \Big( {x_F \over x_F + x_J } ,\bq,\bq + \bk_J \Big)  \,
 D_\Delta\Big({x_+ \over x_F}\Big)  \Big({\bl \over |\bl|} \cdot {\bq \over |\bq|} \Big)  \nonumber \\
&+&  I^f_{\Delta \Delta} \Big( {x_F \over x_F + x_J } ,\bq,\bq + \bk_J \Big) 
D_{\Delta \Delta}\Big({x_+ \over x_F}\Big) \Big[ 2 \Big({\bl \over
  |\bl|} \cdot {\bq \over |\bq|} \Big)^2 - 1 \Big ] \Big\} \; .
\label{eq:x-sec}
\end{eqnarray}

Rapidities are obtained as:
\begin{eqnarray}
 y_i =  \log\Big( {x_i \sqrt{S} \over \sqrt{\bk_i^2} }\Big) \, \leftrightarrow x_i =  \sqrt{ {\bk_i^2 \over S} }\, \cdot \,  e^{y_i}\, , i = +,-,J \, .
\end{eqnarray}

The longitudinal momentum fractions $x_1,x_2$ entering the quark and 
gluon distributions are then
\begin{eqnarray}
 x_1 &=& x_F + x_J = x_+ + x_- + x_J = \sqrt{ \bk_+^2  \over S} e^{y_+}
 + \sqrt{ \bk_-^2 \over S} e^{y_-} + \sqrt{ \bk_J^2 \over S} e^{y_J} \; , \nonumber \\
 x_2 &=& \sqrt{ \bk_+^2 \over S} e^{-y_+} + \sqrt{ \bk_-^2  \over S}
 e^{-y_-} + \sqrt{ \bk_J^2  \over S} e^{-y_J} \; .
\label{x1_x2}
\end{eqnarray}
For completness the invariant mass of the dilepton system is
\begin{eqnarray}
 M^2 = m_{\perp +}^2 + m_{\perp -}^2 + 2 m_{\perp +}m_{\perp-} \cosh(y_+ - y_-) - \bq^2 \, , \, \, \, m_{\perp \pm} = \sqrt{\bk_\pm^2 + m_\pm^2} \, .
\end{eqnarray}

\section{First results}

In the present paper we shall use different UGDFs known from the
literature. The Kimber-Martin-Ryskin distributions \cite{KMR} make
a simple link to collinear distributions. 
In this approach the transverse momentum distribution of ``initial''
gluons originates from the last emission in the ladder. In the present 
calculations we use MSTW08 distributions \cite{MSTW08} to generate 
the KMR unintegrated gluon distributions. Here we use numerical 
implementation by Maciu{\l}a and Szczurek 
used e.g. in the production of charm and double charm
\cite{Maciula-Szczurek}. For the forward emissions considered here
rather low longitudinal momentum fractions enter into the calculations.
In this region a nonlinear effects and onset of saturation may be,
at least potentially, important. The nonlinear effects were
implemented e.g. in Ref.\cite{Kutak:2004ym}. These distributions give a nice
description of forward exclusive production of $J/\psi$ mesons
\cite{Cisek:2014ala}. In addition, for reference, 
we shall use also a simple Golec-Biernat and W\"usthoff
(GBW) parametrization \cite{GBW} and unintegrated gluon distribution
obtained from a dipole-nucleon cross section
solving the Balitsky-Kovchegov equation \cite{Balitsky:1995ub,Kovchegov:1999yj}, 
published in \cite{Albacete:2009fh} which we will name in the present paper AASM UGDF
for brevity. See the appendix for a description of the numerical procedure.

For the quark and antiquark distributions we use MSTW08 leading-order
distributions \cite{MSTW08}. For most of the calculations we used
$M_{ll}^2$ both as a factorization and renormalization scales.
We have also tried:
\begin{eqnarray}
\mu_R^2 &=& \max \left(\kappa_{\perp}^2,q_{\perp}^2+\varepsilon^2 \right) \; , \nonumber 
\\ \mu_F^2 &=& q_{\perp}^2 + \varepsilon^2 \; .
\label{scales}
\end{eqnarray}
The corresponding results turned out to be almost identical.

\subsection{Full rapidity range}

Before going to predictions for particular experiments we wish to
discuss the general situation for the whole phase space, i.e. in the
broad range of lepton rapidities.

\begin{figure}[!ht]
\includegraphics[width=7cm]{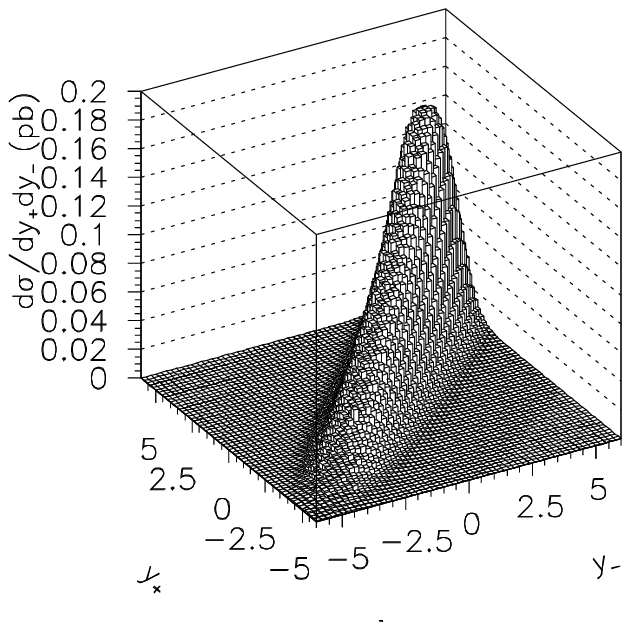}
\includegraphics[width=7cm]{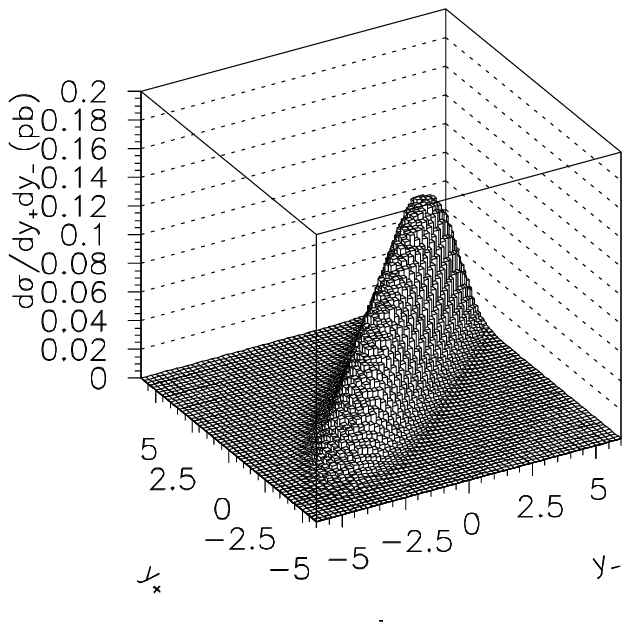}
\caption{\label{fig:full_map_ypym}
Two-dimensional $(y_{+},y_{-})$
distribution for $\sqrt{s}$ = 7 TeV and $k_{T+},k_{T-} >$ 3 GeV
for MSTW08 PDF and KMR (left) and KS (right) UGDFs.
}
\end{figure}

In Fig.\ref{fig:full_map_ypym} we show a two-dimensional distribution
in rapidities of positively and negatively charged leptons. One can observe 
that the contribution of the $(q/\bar q) g \to l^+ l^- (q/\bar q)$ process
extends into a quite broad range, also into the region of negative 
rapidities of positively ($y_{+}$) and negatively ($y_{-}$) charged leptons. 
Similar contribution of the $g (q/\bar q) \to l^+ l^- j$ subprocess
would be trivially symmetric around the $(y_{+}=0,y_{-}=0)$ point. 
The calculation was done with the leading-order MSTW08 quark/antiquark 
distributions and the Kimber-Martin-Ryskin UGDF \cite{KMR}.
The figure clearly shows that including only one of the contributions 
is not sufficient but we wish to stress that this is routinely done 
in the dipole approach (see e.g.\cite{GolecBiernat:2010de,Basso:2015pba}).

\begin{figure}[!ht]
\includegraphics[width=8cm]{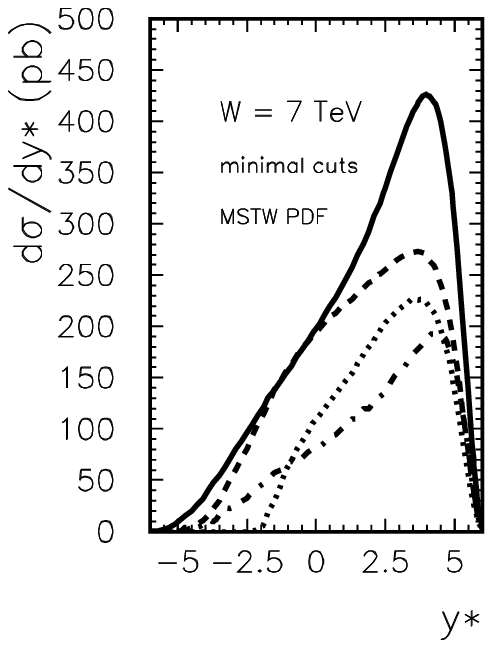}
\caption{\label{fig:full_dsig_dystar_ugdf}
Distribution in rapidity of the dileptons
for $\sqrt{s}$ = 7 TeV and $k_{T+},k_{T-} >$ 3 GeV
for MSTW08 PDF and different UGDFs:
KMR (solid), KS (dashed), AAMS (dotted) and GBW (dash-dotted).
}
\end{figure}

The rapidities of both leptons are strongly correlated i.e.
$y_{+} \approx y_{-}$. Distribution in rapidity of 
the dileptons may be particularly interesting. 
In Fig.\ref{fig:full_dsig_dystar_ugdf} we show
such distributions for different UGDFs from the literature.
Quite different results are obtained for different UGDFs.
It is obvious that at the rapidity of the lepton pair $y_{*} \approx$ 0 
both side mechanisms ($g q/\bar q$ or $q/\bar q g$) must be included. 
At $y_{*} =$0 they give exactly the same contribution.
This is not correctly treated in the dipole approaches where
only one side contribution is included.

\begin{figure}[!ht]
\includegraphics[width=7cm]{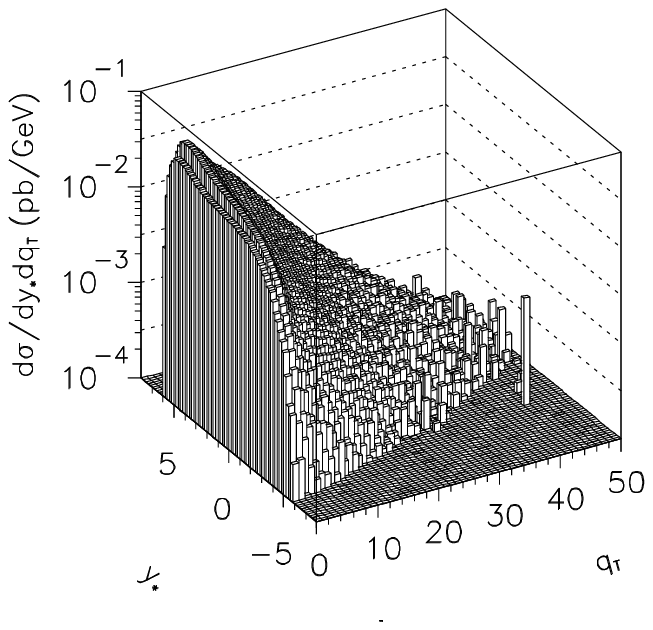}
\includegraphics[width=7cm]{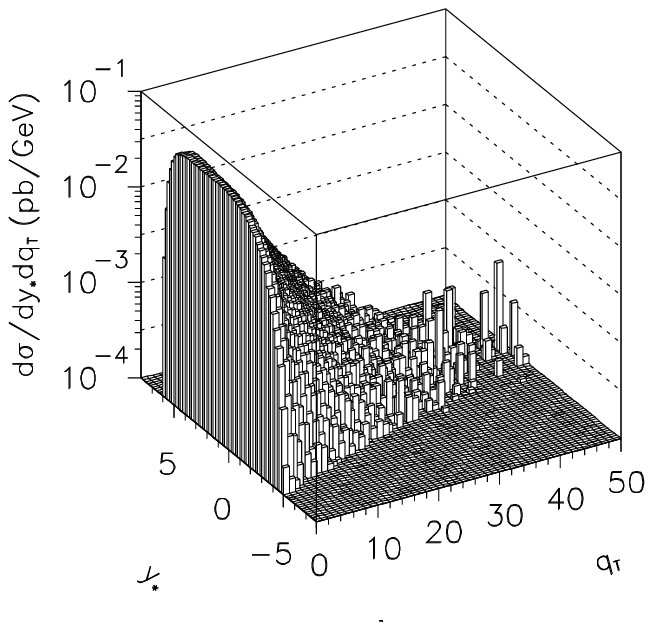}
\caption{\label{fig:full_map_ystarptstar}
Two-dimensional $(y_{*},q_{T})$
distribution for $\sqrt{s}$ = 7 TeV and $k_{T+},k_{T-} >$ 3 GeV
for MSTW08 PDF and KMR (left) and KS (right) UGDF.
}
\end{figure}

In contrast to leading-order collinear approach, in our approach
dileptons have finite transverse momenta.
In Fig.\ref{fig:full_map_ystarptstar} we show two-dimensional
distributions in rapidity and transverse momentum of dileptons.
One can see that at large (positive) rapidities the span of transverse 
momenta is significantly broader. This effect was not discussed so far 
in the literature. In our case the effect is inherently related
to the models of UGDFs used in the calculation. 
Practically all models of UGDFs predict such an effect.
It would be interesting to observe/verify such an effect experimentally 
at the LHC.

\begin{figure}[!ht]
\includegraphics[width=7cm]{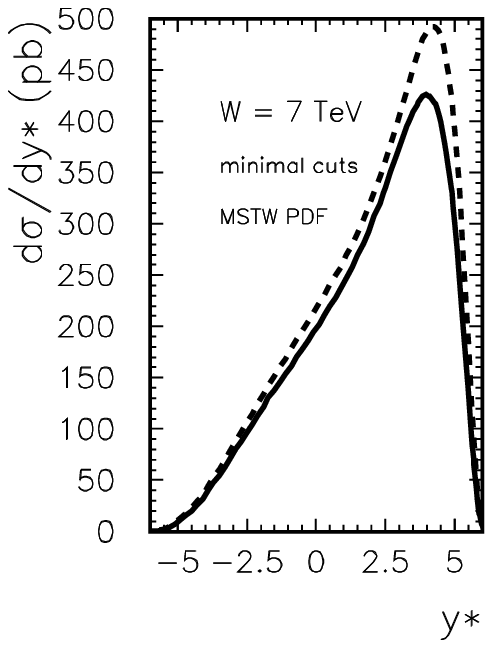}
\includegraphics[width=7cm]{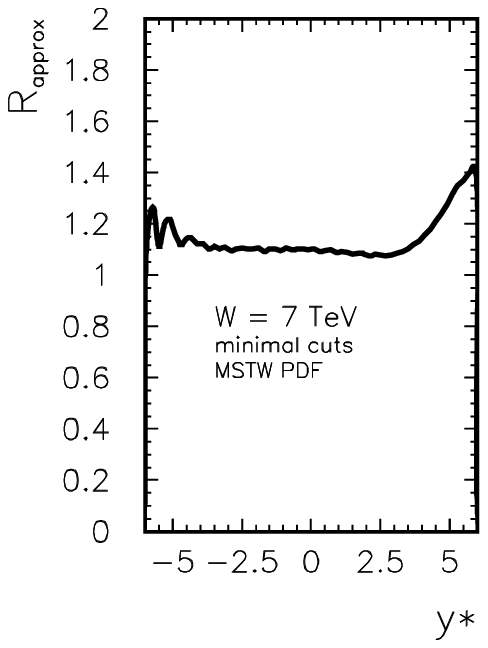}
\caption{\label{fig:full_dsig_dystar_approximatex}
Distribution in $y_{*}$ for exact (solid) and approximate (dashed)
formula for calculating $x_1$ and $x_2$ for $\sqrt{s}$ = 7 TeV and 
$k_{T+},k_{T-} >$ 3 GeV for MSTW08 PDF and KMR UGDF.
In the right panel we show the ratio of the two distributions.
}
\end{figure}

In the traditional dipole approach the produced jet (quark or antiquark)
is not taken into account explicitly into the kinematics of the process. 
In our calculations it enters in the calculation of
parton momentum fractions: $x_1$ (gluon distribution) and 
$x_2$ (quark/antiquark distribution). 
In Fig.\ref{fig:full_dsig_dystar_approximatex} we demonstrate the effect
when the part of $x_i$ corresponding to the jet emission 
(see Eq.(\ref{x1_x2})) is neglected. The largest effect is obtained when 
$y_{*}$ is large i.e. when both charged leptons are produced very forward.
This is also the region when saturation, or more generally nonlinear
effects, are expected. Therefore one should be very careful
in interpreting agreement or disagreement of any calculation in this
region. We shall return to the problem in the context of LHCb
kinematics.  

\begin{figure}[!ht]
\includegraphics[width=8cm]{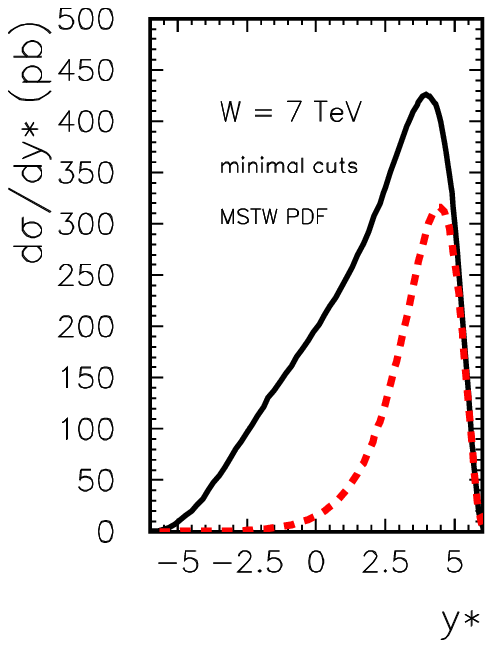}
\caption{\label{fig:full_dsig_dystar_valence}
Distribution in rapidity of the dileptons
for $\sqrt{s}$ = 7 TeV and $k_{T+},k_{T-} >$ 3 GeV
for MSTW08 valence quark distributions and KMR UGDFs.
}
\end{figure}

In the calculations performed so far both valence and sea
quark/antiquark collinear distributions are included.
Fig.\ref{fig:full_dsig_dystar_valence} demonstrates
the role of valence quark distributions (compare the solid (all components)
and the dashed (valence quarks only) lines). 
The contribution related to valence quark distributions is concentrated 
at $y_{*} >$ 0. Only the sea quark/antiquark
contribution extends to the region of $y_{*} <$ 0. This region
is neglected in the most dipole model approaches in the literature.
The LHCb region is dominated by the valence component. We do not need
to mention in this context that the valence quark distributions are
well known and therefore in this region of rapidities one can test
models of UGDFs, provided kinematics of the process is correctly 
taken into account as discussed already above.

\subsection{LHCb}

In this subsection we show results relevant for the LHCb collaboration
results \cite{LHCb-CONF_2012}. The LHCb configuration, due to its specificity
(2.0 $ < \eta <$ 4.5), allows to test very asymmetric longitudinal
momentum fractions of partons. This is potentially interesting
in the context of searches for onset of nonlinear effects and/or
saturation which are expected to occur in the region of very small-$x$ 
of gluons.

\begin{figure}[!ht]
\includegraphics[width=8cm]{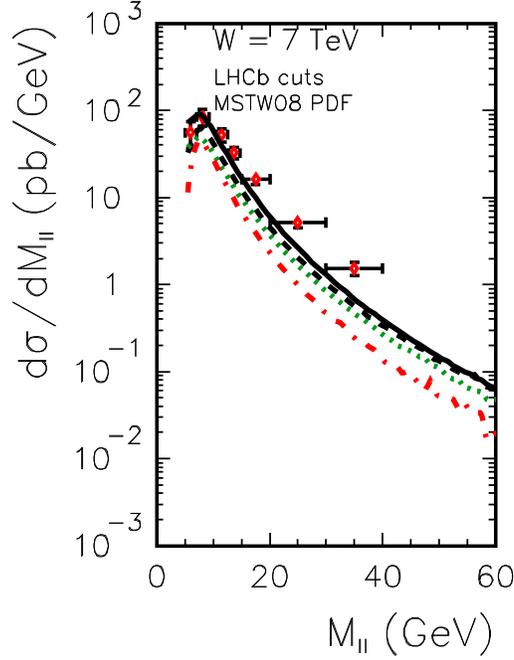}
  \caption{\label{fig:lhcb_dsig_dMll_ugdfs}
Invariant mass distribution (only the dominant component)
for the LHCb cuts: 2 $< y_{+},y_{-} <$ 4.5, $k_{T+},k_{T-} >$ 3 GeV
for different UGDFs: KMR (solid), Kutak-Stasto (dashed), AAMS (dotted)
and GBW (dash-dotted).
}
\end{figure}

Dilepton invariant mass distribution is traditionally the most
popular observable in the context of Drell-Yan processes.
In Fig.\ref{fig:lhcb_dsig_dMll_ugdfs} we show invariant mass
distribution for different UGDFs from the literature.

\begin{figure}[!ht]
\includegraphics[width=5.0cm]{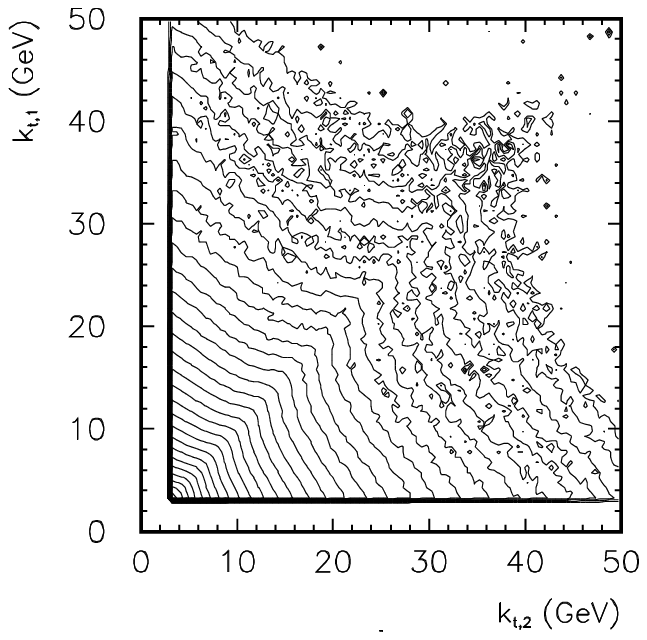}
\includegraphics[width=5.0cm]{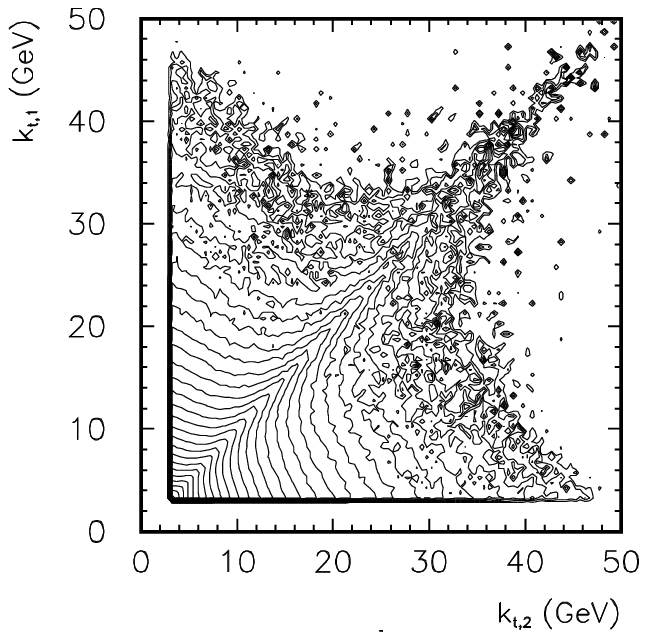}
\includegraphics[width=5.0cm]{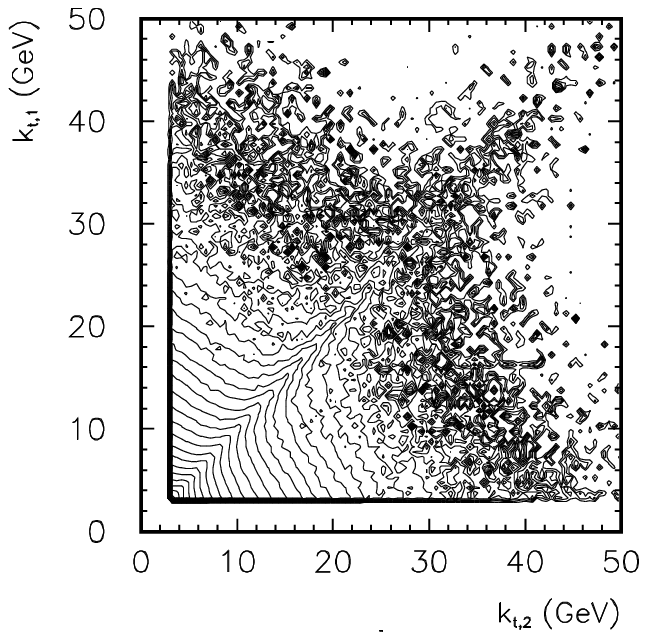}
\caption{\label{fig:lhcb_map_ptpptm}
Two-dimensional $(k_{T+},k_{T-})$
distribution for $\sqrt{s}$ = 7 TeV and $k_{T+},k_{T-} >$ 3 GeV
for MSTW08 PDF and KMR (left), KS (middle) and AAMS (right) UGDFs.
}
\end{figure}

In naive leading-order collinear calculation charged leptons are
produced back-to-back. In the $k_T$-factorization approach presented 
here this is dramatically different. In Fig.\ref{fig:lhcb_map_ptpptm}
we discuss correlations in lepton transverse momenta.
For the KMR UGDF the transverse momenta are much less correlated than
e.g. for the KS or AAMS UGDFs. In the letter cases they are
enhanced for $k_{T+} = k_{T-}$.

The same effect can be demonstrated in one-dimensional distribution
in transverse momentum of the dilepton pairs.
Very different distributions are obtained for different UGDFs.
It would be interesting to compare the results of our calculations with
experimental data.

\begin{figure}[!ht]
\includegraphics[width=8cm]{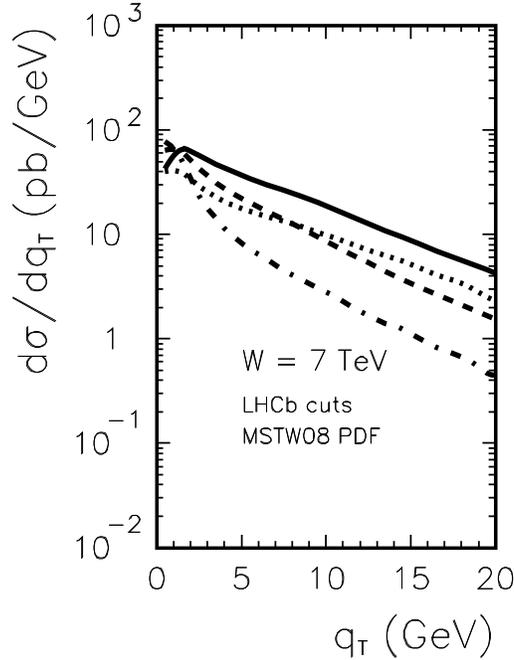}
  \caption{\label{fig:lhcb_dsig_dptpair}
Dilepton transverse momentum distribution (only the dominant component)
for the LHCb cuts: 2 $< y_{+},y_{-} <$ 4.5, $k_{T+},k_{T-} >$ 3 GeV
for different UGDFs: KMR (solid), Kutak-Stasto (dashed), AAMS (dotted)
and GBW (dash-dotted).
}
\end{figure}

\begin{figure}[!ht]
\includegraphics[width=8cm]{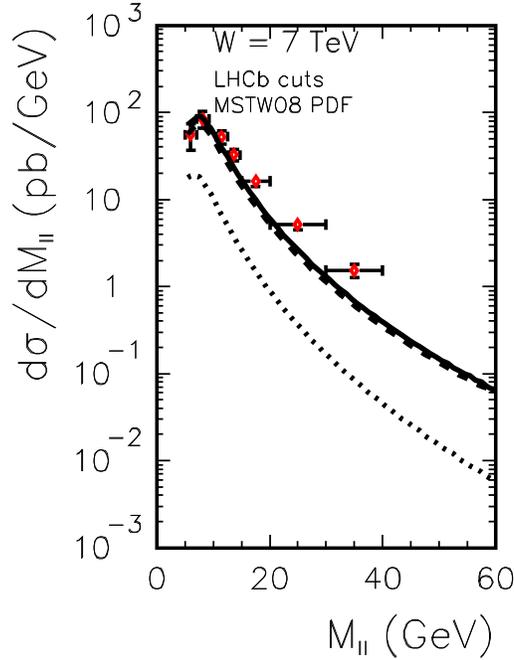}
  \caption{\label{fig:lhcb_dsig_dMll_kmr_LT}
The T and L contributions to the dilepton invariant mass distribution 
for the LHCb kinematics: 2 $< y_{+},y_{-} <$ 4.5, $k_{T+},k_{T-} >$ 3 GeV.
KMR UGDF was used here.
}
\end{figure}

In Fig.\ref{fig:lhcb_dsig_dMll_kmr_LT} we show the invariant 
mass distribution as well as the T and L contributions separately. 
We see that the T contribution is significantly larger than 
the L contribution, especially for large dilepton invariant masses.

\begin{figure}[!ht]
\includegraphics[width=8cm]{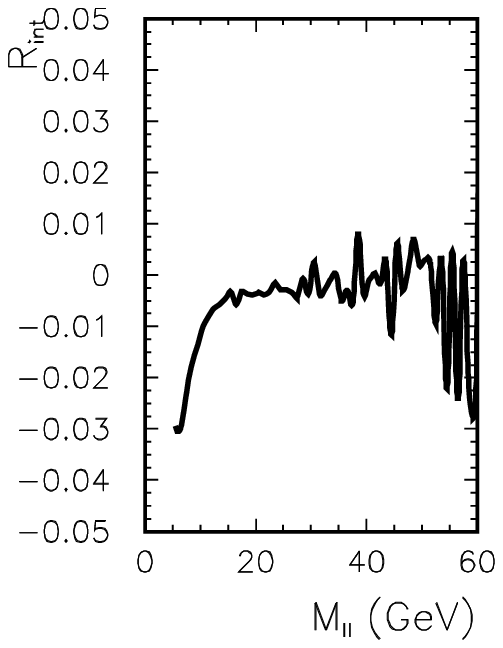}
  \caption{\label{fig:rat_int_Mll}
The $R_{int}$ as a function of $M_{ll}$ for the LHCb kinematics:
2 $< y_{+},y_{-} <$ 4.5, $k_{T+}, k_{T-} >$ 3 GeV.
KMR UGDF was used here. The fluctuations are due to insufficient
statistics of our Monte Carlo calculation.
}
\end{figure}

Now we wish to illustrate the role of the interference terms.
Let us define the quantity:
\begin{equation}
R_{int} = \frac{d \sigma_{all} - d \sigma_{T+L}}{d \sigma_{all}} \; .
\label{relative_interference_effects} 
\end{equation}
As an example in Fig.\ref{fig:rat_int_Mll} we show the so-defined 
quantity as a function of dilepton invariant mass for the LHCb kinematics.
One can observe very small effect of including interference terms of the
order of 1 \%. The fluctuations of the theoretical curve are due to
the Monte Carlo method and smallness of the effect.

\begin{figure}[!ht]
\includegraphics[width=8cm]{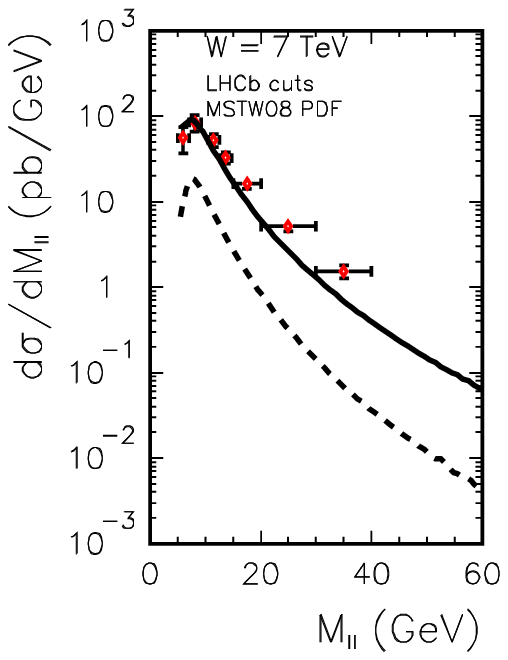}
  \caption{\label{fig:lhcb_dsig_dMll_second}
Contributions of the second-side component for the LHCb kinematics: 
2 $< y_{+},y_{-} <$ 4.5, $k_{T+},k_{T-} >$ 3 GeV. KMR UGDF was used here.
}
\end{figure}

So far we have considered only $g + q/\bar q$ contribution.
Now we wish to discuss how important is the second-side (subdominant)
$q/\bar q + g$ contribution for the LHCb kinematics. In 
Fig.\ref{fig:lhcb_dsig_dMll_second} we show both the dominant
(dashed) and subdominant (dotted) contributions as well as their sum (solid).
Clearly the subdominant contribution is not negligible.

.
\subsection{ATLAS}

In this subsection we show similar results for the low-$M_{ll}$
ATLAS data \cite{Aad:2014qja}. The ATLAS detector covers more central
rapidity range (-2.4 $< y_{+},y_{-} <$ 2.4 ) and imposes a slightly larger
lower cut on the dilepton transverse momenta $k_{T+}, k_{T-} >$ 6 GeV.

\begin{figure}[!ht]
\includegraphics[width=8cm]{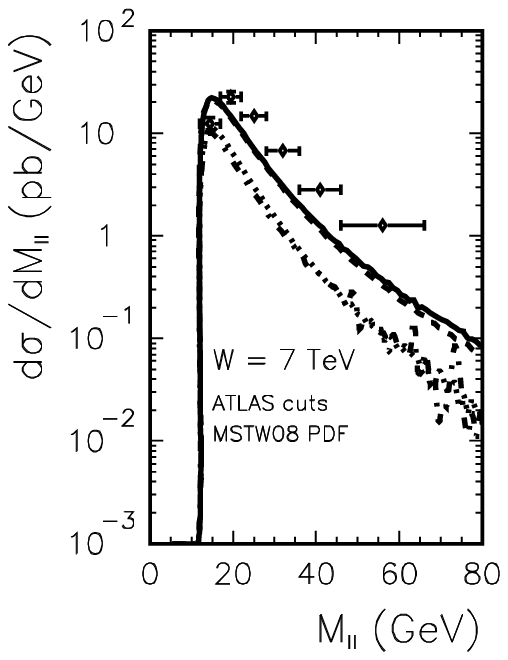}
  \caption{\label{fig:atlas_dsig_dMll}
Invariant dilepton mass distribution for the ATLAS kinematics:
-2.4 $< y_{+},y_{-} <$ 2.4, $k_{T+}, k_{T-} >$ 6 GeV.
Here both $g q/\bar q$ and $q/\bar q g$ contributions have been included.
}
\end{figure}

The invariant mass distribution for the ATLAS kinematics is shown 
in Fig.\ref{fig:atlas_dsig_dMll}. We get relatively good agreement 
with the ATLAS data for dilepton invariant masses $M_{ll}$ at the threshold. 
At larger dilepton invariant mass some strength is clearly missing. 
Here longitudinal momentum fractions are typically $x_1, x_2 \sim$ 0.01-0.1. 
This is a region where antiquark distributions are dominated by the
meson cloud effects (see e.g. \cite{meson_cloud}). 
Some effects of the type of $q \bar q$ annihilation 
are clearly not included in the present approach (as well as in the
dipole approach), at least for the considered range of $x_1$, $x_2$. 

\begin{figure}[!ht]
\includegraphics[width=5cm]{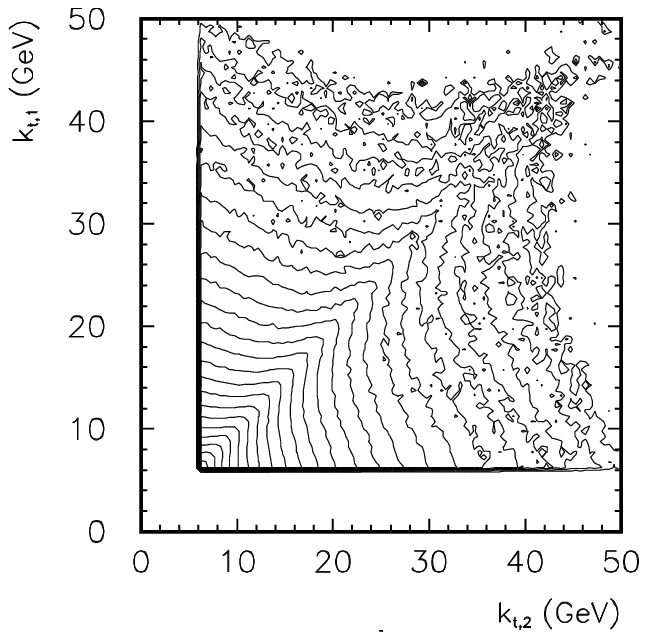}
\includegraphics[width=5cm]{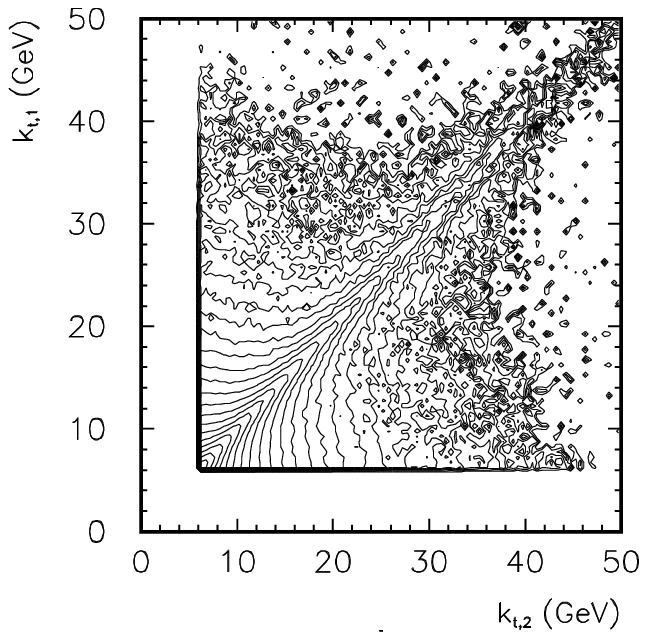}
\includegraphics[width=5cm]{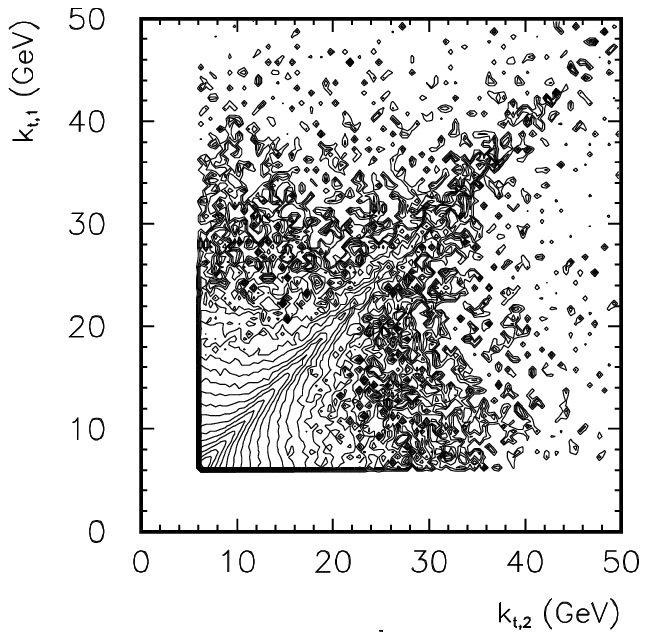}
  \caption{\label{fig:atlas_dsig_dptpdptm}
Lepton transverse momentum correlations for
the ATLAS kinematics: -2.4 $< y_{+}, y_{-} <$ 2.4, $k_{T+}, k_{T-} >$ 6 GeV.
The left panel is for the KMR UGDF, the middle panel for the KS UGDF
and the right panel for the AAMS UGDF.
}
\end{figure}

The transverse momenta of leptons are correlated as shown in Fig.
\ref{fig:atlas_dsig_dptpdptm}. We observe a clear ridge along
$k_{T+} = k_{T-}$ and enhancements when either $k_{T+}$ or $k_{T-}$
are small.

\begin{figure}[!ht]
\includegraphics[width=8cm]{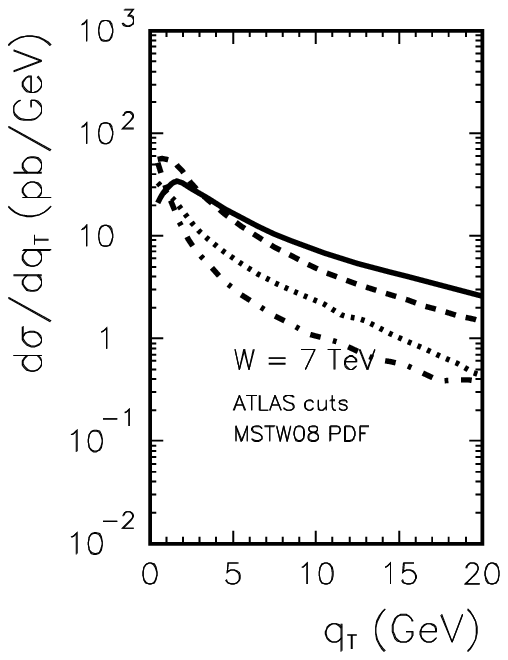}
  \caption{\label{fig:atlas_dsig_dptpair}
Transverse momentum distribution of dileptons for the
ATLAS kinematics: -2.4 $< y_{+},y_{-} <$ 2.4, $k_{T+}, k_{T-} >$ 6 GeV for MWST08 PDF and
for different UGDFs:
KMR (solid), KS (dashed), AAMS (dotted) and GBW (dash-dotted). 
}
\end{figure}

Distributions in transverse momentum of the dilepton pairs
are shown in Fig.\ref{fig:atlas_dsig_dptpair} for the different UGDFs.
This plot reminds corresponding plot for the LHCb kinematics.

\section{Summary and conclusions}

In the present paper we have considered Drell-Yan production of
dileptons in the forward rapidity region in a hybrid high-energy
approach.
In this approach the main ingredients are collinear quark/antiquark 
distributions and unintegrated gluon distributions.
Corresponding formula for matrix element in high-energy
approximation has been derived and presented.
The relation to the commonly used dipole approach has been discussed.
In contrast to the dipole approach our formulation correctly
treats the kinematic of the process and can be applied to the analysis
of real experimental data including their specific kinematic cuts.
In our more general formula we have obtained four terms instead of two 
(T, L) in the standard dipole model.

A corresponding program including underlying $2 \to 3$ subprocess
matrix elements ($g + q/\bar q \to l^+ l^- j$ or 
                 $q/\bar q + g \to l^+ l^- j$), 
PDFs and UGDFs has been constructed.
To illustrate our approach we have performed calculations of 
differential cross sections corresponding to recent experimental results
for low-mass dilepton production by the LHCb and ATLAS collaborations.
In the first calculation we have used different UGDFs from the
literature and MSTW08 quark/antiquark distributions.
Relatively good agreement with the experimental
data has been achieved for small $M_{ll}$. Some strength at larger
$M_{ll}$ is missing which is probably due to lack of meson cloud
effects, not included here.

In contrast what was done in the literature, we have found
that both side contributions have to be included even for
the LHCb configuration. For the ATLAS kinematics this gives half
of the cross section.

We have found that the contribution of individual terms (i = T, L, ...)
strongly depends on kinematical variables (such as $M_{ll}$) 
as well as on cuts.
We have quantified the effect of the new interference terms not present 
explicitly in the dipole approach.
We have found that the missing strength at larger $M_{ll}$ could be 
due to e.g. meson cloud effects and the perturbative gluon component
alone considered here may be not sufficient.

We do not see clear hints of saturation at small $M_{ll}$.
We wish to stress also that this region of the phase space is potentially
difficult for extracting the Drell-Yan contribution due to
potential contamination of double semi-leptonic decays
of charmed and/or bottom mesons or baryons which is slightly model 
(Monte Carlo) dependent.

\section*{Acknowledgments}

We are indebted to Rafal Maciula for help in adopting our program
to a Monte Carlo form.
This study was partially supported by the Polish National Science Centre 
grant DEC-2014/15/B/ST2/02528.

\appendix
\section{Unintegrated gluon distribution from dipole cross sections}

The dipole cross section is related to the unintegrated glue as
\begin{eqnarray}
 \sigma(x,\br) = {4 \pi \over N_c} \int {d^2\bkappa \over \bkappa^4} \alpha_S {\cal{F}}(x,\bkappa) \Big\{ 1 - \exp(i \bkappa \br) \Big\} \, .
\end{eqnarray}
The parametrizations of \cite{Albacete:2009fh} are presented in the form
\begin{eqnarray}
 \sigma(x,\br) = \sigma_0 \cdot N(x,\br) \, ,
\end{eqnarray}
with $N(x,\br) \to 1$ at large $\br$.
We can therefore easily obtain, that 
\begin{eqnarray}
 {\alpha_S {\cal{F}}(x,\bkappa) \over \bkappa^4} = {\sigma_0 N_c \over 4 \pi} \int {d^2\br \over (2 \pi)^2}  \exp(-i \bkappa \br) \Big[ 1 - N(x,\br) \Big] \, ,
\end{eqnarray}
or
\begin{eqnarray}
 {\cal{F}}(x,\bkappa) = {\sigma_0 N_c \over 8 \pi^2} { \bkappa^2 \over \alpha_S(\bkappa^2)} \int_0^\infty r dr J_0(\kappa r) \Big[ 1 - N(x,\br)\Big] ,
\label{eq:FB_trf}
\end{eqnarray}
where $J_0(x)$ is the Bessel function. The Fourier-Bessel (or Hankel-) transform (\ref{eq:FB_trf}) can pose severe numerical problems, if
values at large $\bkappa^2$ are required.
For the evaluations of these integrals we use therefore a dedicated code FFTLog \cite{Hamilton:1999uv} which is based on
the algorithm of \cite{Talman}.

\end{document}